\documentclass{mn2e}
\usepackage[normalem]{ulem}
\newcommand{\apropto}{\;
  \raise0.3ex\hbox{$\propto$\kern-0.75em\raise-1.1ex\hbox{$\sim$
  }}\;\hskip-2pt }
\usepackage{graphicx}
\usepackage{subfigure}

\newcommand{\lta}{\;
  \raise0.3ex\hbox{$<$\kern-0.75em\raise-1.1ex\hbox{$\sim$
  }}\;\hskip-2pt }
\newcommand{\gta}{\;
  \raise0.3ex\hbox{$>$\kern-0.75em\raise-1.1ex\hbox{$\sim$
  }}\;\hskip-2pt }

\title[Solar small-scale dynamo and polarity of sunspot groups
]{Solar small-scale dynamo and polarity of sunspot groups
 }

\author[D.~Sokoloff, A.~Khlystova, V.~Abramenko]
{D.~Sokoloff$^{1}$\thanks{email: sokoloff.dd@gmail.com},
A.~Khlystova$^{2}$\thanks{email: hlystova@iszf.irk.ru},
V.~Abramenko$^{3}$\thanks{email: vabramenko@gmail.com}\\
$^{1}${Department of Physics, Moscow State University, Moscow, 119992, Russia}\\
$^{2}${The Institute of Solar-Terrestrial Physics, Siberian Branch, Russian Academy of Sciences, Irkutsk, 664033, Russia}\\
$^{3}${The Central Astronomical Observatory of the Russian Academy of Science, Saint-Petersburg, 196140, Russia}\\
}
\begin{document}
\label{firstpage} \maketitle
\begin{abstract}
In order to clarify a possible role of small-scale dynamo in formation of solar 
magnetic field, we suggest an observational test for small-scale dynamo action 
based on statistics of anti-Hale sunspot groups. As we have shown, according to 
theoretical expectations the small-scale dynamo action has to provide a 
population of sunspot groups which do not follow the Hale polarity law, and the 
density of such groups on the time-latitude diagram is expected to be 
independent on the phase of the solar cycle. Correspondingly, a percentage of 
the anti-Hale groups is expected to reach its maximum values during solar 
minima. For several solar cycles, we considered statistics of anti-Hale groups 
obtained by several scientific teams, including ours, to find that the 
percentage of anti-Hale groups becomes indeed maximal during a solar minimum. 
Our interpretation is that this fact may be explained by the small-scale dynamo 
action inside the solar convective zone.

\end{abstract}

\section{Introduction}

Solar magnetic activity is widely believed to be associated with dynamo action 
somewhere in the solar convective shell. Identification of particular details of 
solar dynamo with surface manifestations of solar dynamo accessible for 
observations remains however a disputable problem. The point is that many 
important details of dynamo action being hidden in solar interior can not be 
observed directly, and we have to learn about them basing on indirect tracers. 
The link between a particular parameter important for a solar dynamo model and 
observable tracers of solar activity may be quite complicated, so adding new 
tests for comparison between concepts of solar dynamo and observations is a very 
attractive and simultaneously highly nontrivial undertaking.

Recently a new physical entity -- solar small-scale magnetic field -- was 
suggested for observational verification of solar dynamo concepts. Solar 
magnetic field obviously contains some small-scale details which hardly can be 
included in the global solar magnetic field produced by traditional mean-field 
dynamo models. Of course, more detailed dynamo models based on direct numerical 
simulations of non-averaged MHD-equations give dynamo-driven magnetic 
configurations much more complicated than those produced by the mean-field 
models. This fact is in agreement with theoretical expectations from mean-field 
models because certain terms in mean-field equations appear as a result of 
averaging of magnetic fluctuations.

The point however is that, apart from the mean-field (or global) dynamo based on 
a joint action of solar differential rotation and mirror-asymmetric convection, 
the dynamo theory (in the framework of average description in terms of 
correlation tensor, see, e.g., a review by Zeldovich et al. (1990), as well as 
in direct numerical simulations, see, e.g., a review by Brandenburg et al. 
(2012) predicts an additional mechanism of magnetic field self-excitation, 
so-called small-scale dynamo which produces small-scale magnetic field, i.e., 
magnetic fluctuations. In other words, the dynamo theory suggests that the total 
dynamo-driven magnetic field $\bf H$ can contain the following contributions: 
mean magnetic field $\bf B$, small-scale magnetic field $\bf b$ connected with 
the mean field $\bf B$, and small-scale magnetic field $\bf h$ generated 
independently of $\bf B$. Theoretical distinction between $\bf b$ and $\bf h$ 
was known quite a long time ago (see, e.g., Brandenburg \& Subramanian, 2005). 
However a general presumption was that it is more or less hopeless to 
distinguish between $\bf b$ and $\bf h$ observationally. As a result, the 
relationship between $\bf b$ and $\bf h$ was very rare addressed in dynamo 
theories. A naive theoretical expectation was that both sources of small-scale 
magnetic field contribute somehow in magnetic fluctuations (as soon as 
fluctuations are ubiquitous in cosmic phenomena), however any attempt to 
distinguish between them hardly can be a fruitful undertaking.

Modern progress in solar magnetic field observations (e.g., Ishikawa et al., 
2007, Lites et al., 2008, a review by de Wijn et al., 2009, Abramenko et al., 
2010, Martinez Gonzalez et al., 2012, to mention a few) makes it possible to 
explore in detail various solar small-scale magnetic structures which often 
look quite specific in comparison with mean solar magnetic field. A natural 
naive expectation appears that we clearly see imprints of the small-scale dynamo 
action, i.e. the field $\bf h$. The fundamental conceptual progress here occurs 
due to the papers by Stenflo (2012, 2013). He critically examined "the relative 
contributions of these two qualitatively different dynamos to the small-scale 
magnetic flux" with the following conclusion: "The local dynamo does not play a 
significant role at any of the spatially resolved scales, nearly all the 
small-scale flux, including the flux revealed by Hinode, is supplied by the 
global dynamo".

We appreciate the importance of the clear and convincing statement of the 
problem formulated by Stenflo (2012, 2013). The point however is that, generally 
speaking, the small-scale dynamo can act in the solar interior and contribute 
very little into surface observables. On the contrary, certain versions of the 
dynamo theory (e.g., Cattaneo \& Tobias, 2014) predict that "large-scale dynamo 
action can only be observed if there is a mechanism that suppresses the 
small-scale fluctuations". So, we deal with a fundamental physical problem and 
it looks reasonable to spend efforts in order to find tiny surface spores of the 
small-scale dynamo action in the solar interior.

In this paper, we suggest that an imprint of the small-scale dynamo action in 
solar interior can be hidden in statistics of sunspot groups which violate the 
Hale polarity law. The law states that in odd, for example, cycles, bipolar 
groups in the Northern (Southern) hemisphere have a positive (negative) magnetic 
polarity of the leading sunspot. We refer to the groups, which violate this law 
as anti-Hale groups. The key issue of the suggested test is that observations 
(as well as direct numerical simulations) deal with the total magnetic field 
$\bf H$ while the underlying physical problem is formulated in terms of the 
statistical quantities originated from the mean-field approach. And it is far 
from straightforward how to make inferences about the later in terms of the 
observed $\bf H$. In other words, we need an explicit description of how we 
separate various contributions to the total magnetic field. In statistical 
studies, such a description is referred as a probabilistic model.

\section{Probabilistic model for the magnetic field}

The start point of the observational test suggested is as follows.

It is well-known that through a solar cycle, the polarity of sunspot groups 
follows the so-called Hale polarity law, i.e., leading sunspots have opposite 
polarities in Northern and Southern hemispheres. Besides, from cycle to cycle, 
the leading-spot polarity changes. The Hale polarity rule reflects symmetry of 
the mean solar magnetic field. We recall that in the framework of dynamo studies 
the number of sunspot groups is believed to be related to the mean field 
strength.

Of course, there are few sunspot groups which do not follow the Hale polarity 
rule. Suppose that anti-Hale groups can be considered as a result of some 
magnetic fluctuations which break symmetry of the mean magnetic field. Suppose 
also that the amplitude of fluctuations is governed by the mean field strength, 
as it should be if the small-scale dynamo does not work. Then we have to expect 
that the number of sunspot groups which follow the Hale polarity law should be 
proportional to the number of anti-Hale groups, and the relative number (the 
percentage) of the anti-Hale groups should be cycle independent.

In contrast, if there is a substantial source of magnetic fluctuations 
independent on the mean field, then the number of anti-Hale groups has to 
contain a cycle-independent component. In this case, the relative number of 
anti-Hale groups has to be enhanced during each solar minimum. This expectation 
admits an observational verification (at least in principle) and looks 
reasonable at least in the first sight. The point however is that realization of 
the above scheme is not a trivial task. Below we present the key features of our 
approach in the most simple form as a toy model of relationship between the 
sunspots polarity and two contributors to magnetic fluctuations.

Let us consider a given solar hemisphere in a given solar cycle and assume that 
according to the Hale polarity law toroidal component of the mean magnetic field 
(which determines polarity of sunspot groups) is directed, say, westwards. Let 
the magnetic field be organized in tubes of two types: the first type are tubes 
produced by the mean field, and the second type tubes are associated with the 
small-scale dynamo action. For the sake of simplicity, we adopt that the tubes 
are directed {\bf longitudinally}, i.e., oriented along the equator.

Let the tubes of the first type contain the magnetic field of two components. 
The fist component, $\bar {\bf B}$, is a non-random flux directed westwards. The 
number density of the tubes is denoted as $n_1$, so that $n_1 \bar {\bf B}$ is 
the mean field $\bf B$. The number density $n_1$ is modulated by 11-year cycle, 
which results in the 11-year modulation of $\bf B$. At the same time, $\bar {\bf 
B}$ is time-independent. We presume that sunspot formation is a threshold 
phenomenon with a threshold magnetic field strength, $\hat B$, slightly 
exceeding the magnitude of $\bar {\bf B}$.

The other magnetic field component is a random (say, Gaussian) magnetic field 
$\bf b$ which is directed {\bf longitudinally} as well, however its mean value 
vanishes, and the field is directed with the equal probabilities west- or 
eastwards. The r.m.s. value of this field, $\bar b$, is proportional to $\bar 
B$. If $\bf b$ is directed westwards and $\bar {\bf B} + {\bf b}$ exceeds $\hat 
B$, then a magnetic tube arises on the solar surface and creates a sunspot group 
which follows the Hale polarity law. If, however, the field $\bf b$ is directed 
eastwards and, by chance, it is so strong that $|\bf b| - |\bar {\bf B}|$ 
exceeds $\hat B$, then the tube arises as well, but the newcomer violates the 
Hale polarity law. A simple estimates (Khlystova \& Sokoloff, 2009; Sokoloff \& 
Khlystova, 2010) show that if $\bar b/\bar B \approx 1$, the number of anti-Hale 
groups should be a few percent of the total number of sunspot groups.

Note that in order to explain why sunspots arise at solar minima (when the mean 
magnetic field is low), we have to suppose that $\bar B$ is time-independent and 
the mean field cyclic modulation comes from the modulation of $n_1$.

The tubes of the second type contain a random (say, Gaussian) magnetic field 
$\bf h$ with a cycle-independent r.m.s. value $\bar h$ and zero mean value. They 
do not contribute to the mean magnetic field. When $|\bf h|$ exceeds $|\hat B|$, 
a tube arises, and the polarity orientation depends on the occasional tube's 
orientation: a Hale polarity law group appears if $\bf h$ was directed westward, 
and an anti-Hale group appears if $\bf h$ was directed eastward.

Obviously, the {\it number} of anti-Hale groups originated from the first-type 
tubes depends on the cycle (as soon as the total number of sunspots diminishes 
at solar minima), while that originated from the second-type tubes is 
cycle-independent. On the contrary, the {\it percentage} of anti-Hale groups 
originated from the first-type tubes does not depend on the cycle, while the 
percentage of that produced by the second-type tubes should reach its maximum 
value at the solar minima because the total number of sunspots is lowered at 
that times. Thus, exploring a presence/absence of cyclic modulation of the 
anti-Hale groups, we can shed light on a relationship between the small-scale 
field $\bf h$, originated from the small-scale dynamo, and the small-scale field 
$\bf b$, produced by the mean-field dynamo action.

The above model is an obvious oversimplification. We only include the features 
that are ultimately required to illustrate our observational test. Further steps 
will be to admit that the magnetic tubes might be directed not exactly 
{\bf longitudinally}, to include temporal growth of magnetic field in a given tube due 
to the dynamo action, etc.

\begin{figure}
\includegraphics[width=0.50\textwidth]{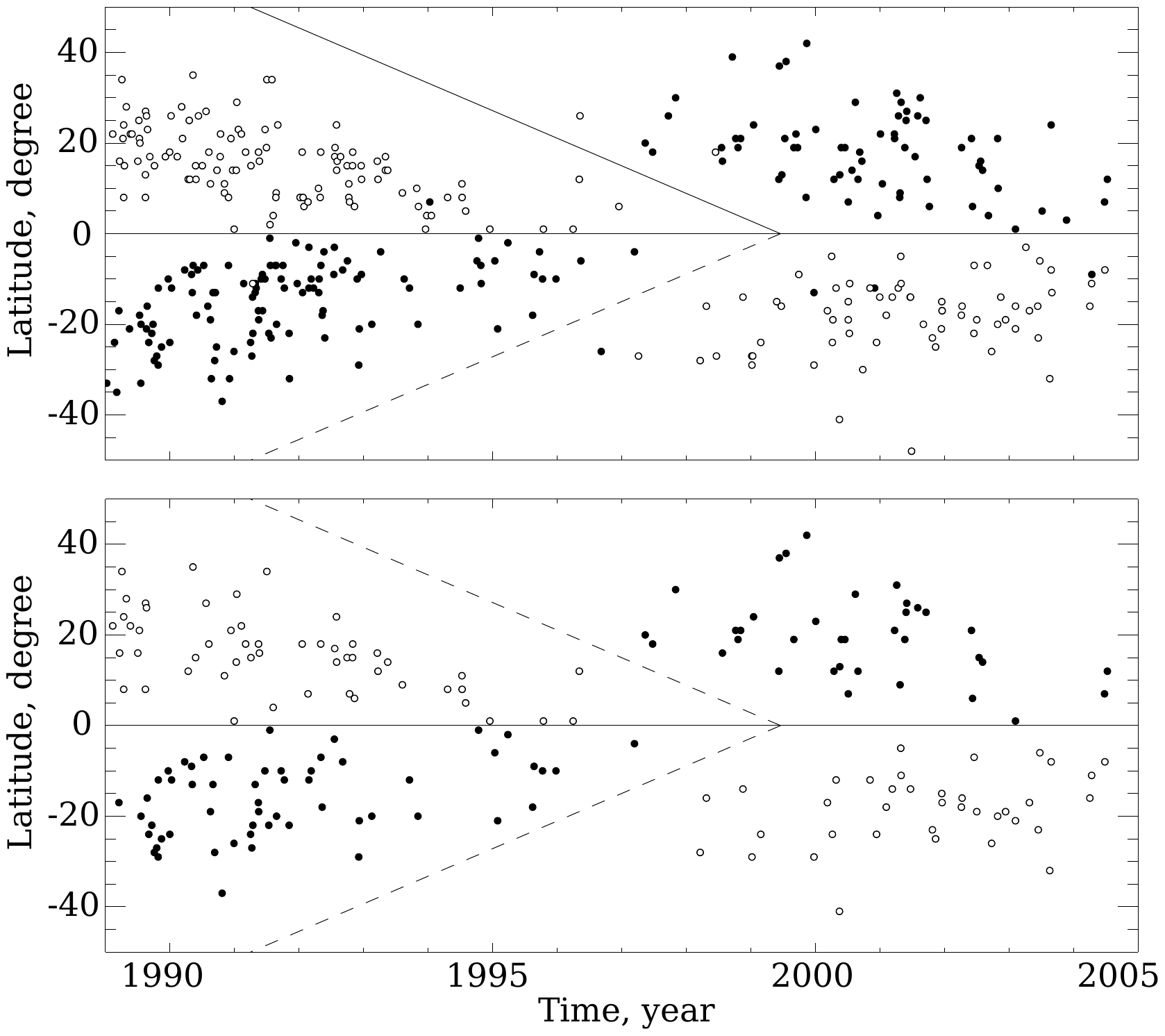}
\caption{Time-latitude distribution anti-Hale sunspot groups for 1989--2004. 
Top: all sunspot groups which were marked as anti-Hale in WSO catalog. Bottom: 
anti-Hale sunspot groups after our checking. White (black) dots indicate groups 
with the positive (negative) magnetic field in the leading sunspots. Dotted 
lines correspond to the boundaries between 22 and 23 solar cycles in accordance 
with method of McClintock et al. (2014).}
 \label{Rich}
\end{figure}

\section{Data}

The Mount Wilson Observatory (MWO) has carried out magnetic observations of 
sunspot groups throughout a century. On the basis of those observations, the 
catalog of magnetic classes of sunspot groups was compiled, which contains the 
information on number of anti-Hale sunspot groups within certain time intervals. 
Unfortunately, the digital format of the complete MWO catalog does not exist 
yet.

Monthly catalogs for 1920--1958, which include the magnetic classification of 
sunspot groups, had been published in each issue of the \textit{Publications of 
the Astronomical Society of the 
Pacific}\footnote{http://www.jstor.org/page/journal/publastrsocipaci/about.html} 
starting with a paper titled: "Summary of Mount Wilson Magnetic Observations of 
Sun-Spots for May and June" (1920). From 1962 to 2009, the catalog data had been 
published in the \textit{Solar-Geophysical Data}, however there were gaps in the 
information on anti-Hale sunspot groups. There is a digital MWO catalog covering 
the time interval from 1962 to 2004. This one is available at the website of the 
\textit{National Geophysical Data 
Center}\footnote{ftp://ftp.ngdc.noaa.gov/STP/SOLAR\_DATA/ 
SUNSPOT\_REGIONS/Mt\_Wilson}, however the information on anti-Hale groups is 
presented for the 1989--2004 time interval only. 

Note that in the MWO catalog, normal and anti-Hale sunspot groups are determined 
by the sign of magnetic field in the leading sunspots and by the group's tilt 
relative to the E--W direction. This method works well for a majority of bipolar 
groups. However, uncertainties can appear in some cases. Thus, the tilt of the 
sunspot axis can be misaligned with the tilt of magnetic polarities, and as a 
result, a normal group with a rare unusual tilt can be classified as an 
anti-Hale one; groups of a singular sunspot cannot be classified; besides, in 
some cases, the identification of a group as an anti-Hale one is ambiguous 
because the group can change the polarity orientation during its evolution. To 
frame a group from its neighbors sometimes is not a unique procedure (see, for 
example, the magnetic complex AR NOAA 09393/09394). We acknowledge the above 
shortcomings, however we decided to use the catalog data.

Richardson (1948) studied in detail anti-Hale sunspot groups from 1917 to 1945 
using the MWO catalog data published in Hale \& Nicholson (1938) and in 
\textit{Publications of the Astronomical Society of the 
Pacific}\footnotemark[1]. He examined the anti-Hale sunspot groups on the 
original records. Table 5 in his paper gives yearly numbers of normal and 
anti-Hale sunspot groups in the Northern and Southern hemispheres separately for 
the 15th, 16th and 17th solar cycles. Richardson normalized the data: the number 
of sunspot groups observed at MWO during a year was multiplied by 365 and 
divided by the number of observation days in a year, which resulted in 
non-integer numbers of anti-Hale groups in his Table 5. The data from this table 
for 1917--1945 were used in the present study.

During periods of cycles overlapping, the high-latitude groups are thought to 
belong to new cycle, and therefore, they have the opposite (to the old cycle) 
magnetic polarity of the leading sunspots. These groups must be marked as 
normal, not anti-Hale ones. Having this in mind, Richardson introduced the cycle 
separation of groups, so that during the cycles overlapping, the low-latitude 
groups belong to the old cycle, whereas the high-latitude groups belong to the 
new cycle. The mark "anti-Hale" was attributed depending on the cycle. Figure 1 
of Richardson (1948) demonstrates this for anti-Hale sunspot groups. We follow 
this rule in the present study.

To obtain the annual number of normal (anti-Hale) groups from Table 5 from 
Richardson (1948), we combined the normal (anti-Hale) groups of both 
hemispheres. Besides, for years of cycles overlapping (1922--1924, 1933--1935, 
1943--1945), we combined low-latitude normal (anti-Hale) groups of old cycle 
with high-latitude normal (anti-Hale) groups of new cycle. As a result of these 
efforts, our first data set, named hereinafter as the 1917--1945 data set, was 
compiled.

Our second data set was compiled on the basis of the above mentioned 1989--2004 
MWO data. In detail, these data were observed from January 1, 1989 to August 
31, 2004. They are available at the website of the \textit{National Geophysical 
Data Center}\footnotemark[2]. There, for anti-Hale groups, the sign "+" is added 
to the magnetic class, according to the catalog 
description\footnote{http://obs.astro.ucla.edu/spotlgnd.html}. Total 354 groups 
were marked as anti-Hale groups. (Note, that during the time period under study, 
the data have gaps, namely, November 1995, December 1998, and January to March 
2002.) We scrutinized each (out of 354) group marked as "anti-Hale". Using the 
data about the magnetic polarity in 
sunspots\footnote{http://obs.astro.ucla.edu/cur\_drw.html}, we determined the 
polarity of leading spots. The time-latitude diagram was compiled (Figure 1, 
top frame). Following the method used by McClintock et al. (2014), we separated 
wings of activity waves to determine the boundaries between cycles (dashes lines 
in Figure 1). The diagram shows that there are certain mistakes in attributing 
the "anti-Hale" property (e.g., black dots are present in the midst of white, 
and visa versa; two high-altitude sunspot groups of the new cycle were wrongly 
marked as anti-Hale groups in 1996). This forced us to re-examine all groups on 
the diagram by means of full disk magnetograms of different observatories 
available at the DPD website\footnote{http://fenyi.solarobs.unideb.hu/DPD/} 
(Gy\H{o}ri et al., 2011). We found that about one-half of identifications were 
erroneous. The mis-identifications are associated predominantly with long-lived 
spots in decaying active regions with formation of small spots or pores of 
opposite polarity to the west from the main spot, or mis-identifications of 
boundaries of sunspot groups in decaying activity complexes. The corrected data 
set 1989--2004 was used to explore the anti-Hale statistics in the next section 
(Figure 1, bottom frame).

In our study, we also consider the results of McClintock et al. (2014). The 
authors analyzed magnetic tilt angles of sunspot groups along with information 
on the magnetic polarities of leading spots. Their statistics of anti-Hale 
bipolar sunspot regions for 1974-2012 was obtained from the data of Li \& Ulrich 
(2012), who determined sunspot magnetic tilt angles from MWO sunspot records and 
daily averaged magnetograms, as well as from SOHO/MDI magnetograms.

\section{Cyclic variations in the percentage of the anti-Hale groups}

The results for the 1917--1945 data set are shown in Figure 2. The upper panel 
shows the annual average of sunspot groups and demonstrates 11-year solar 
activity cycle. The lower panel represents the annual average percentage of 
anti-Hale groups. The percentage of anti-Hale groups tends to be enhanced during 
the cycles' minima. The similar result was obtained for the 1989--2004 data set, 
Figure 3. We found that the observed enhancement of the percentage becomes less 
pronounced when we average data over time intervals longer than 1 year. On the 
contrary, averaging over shorter time intervals seem to be interesting, and it 
is considered below.

\begin{figure}
\includegraphics[width=0.50\textwidth]{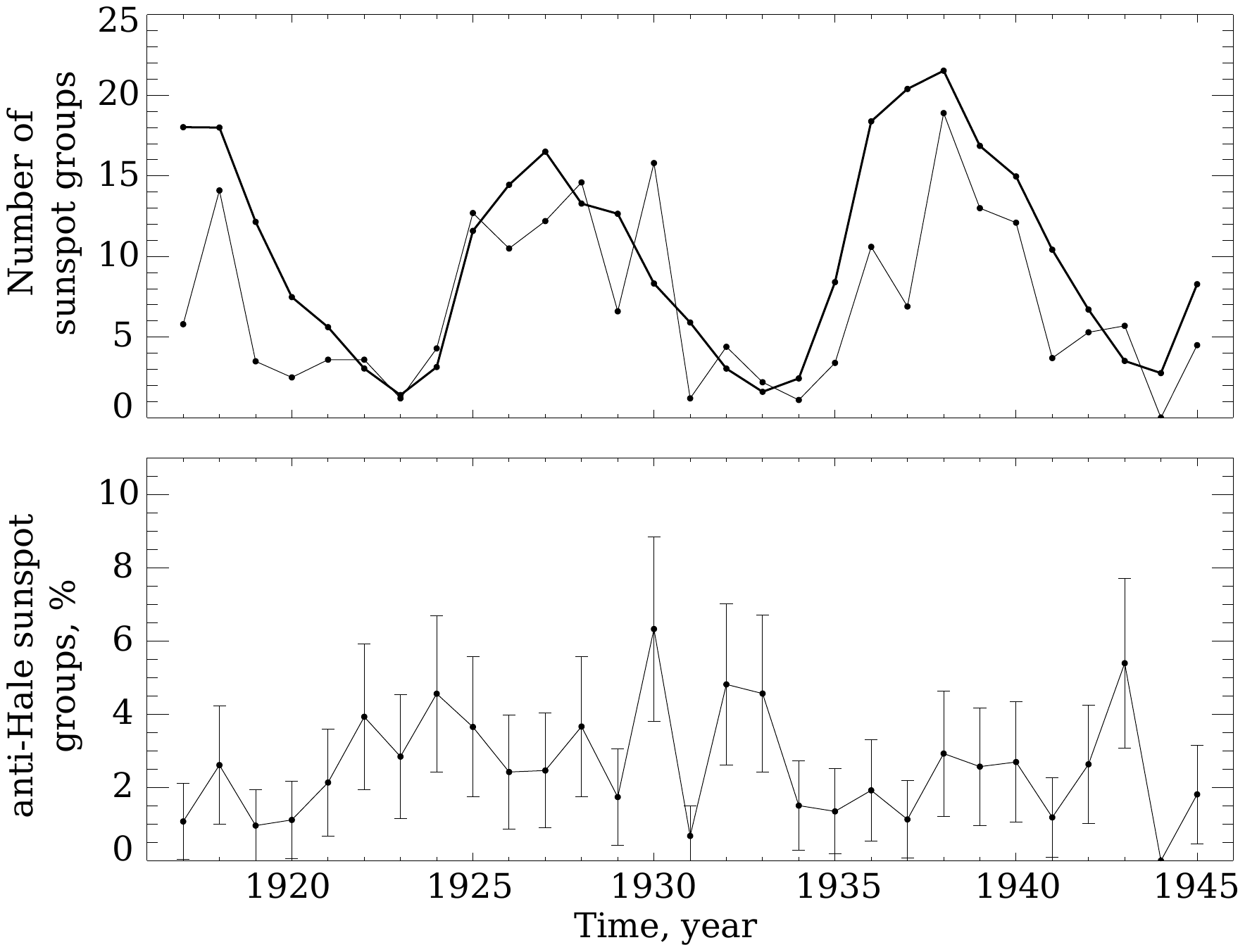}
\caption{Annual average sunspot groups data for 1917--1945. Top: time variations 
of the total number of sunspot groups divided by 30 (thick line) and the 
numbers of anti-Hale groups (thin line). Bottom: the percentage of anti-Hale 
groups. Error bars show standard deviations.}
 \label{Rich}
\end{figure}

\begin{figure}
\includegraphics[width=0.50\textwidth]{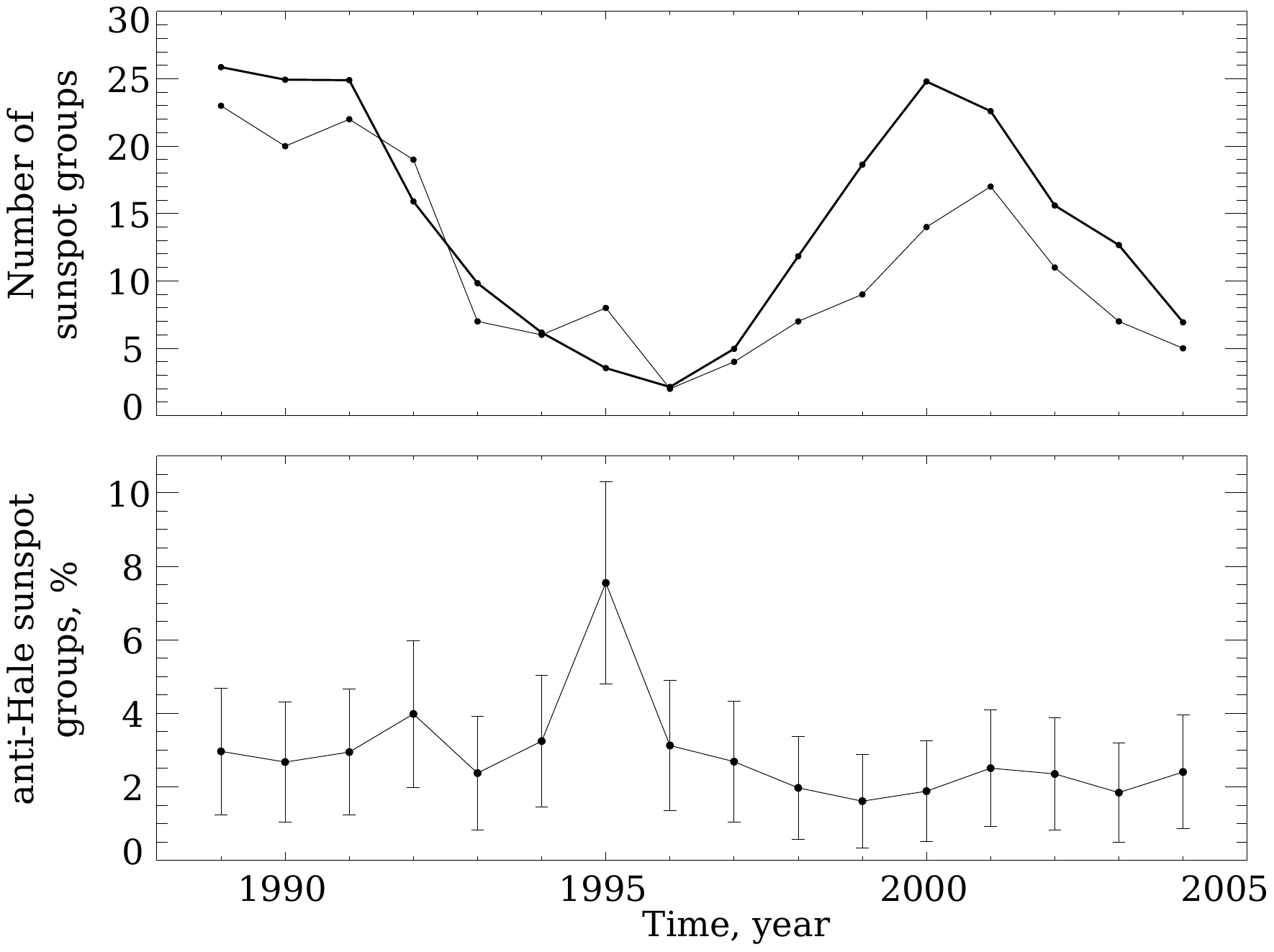}
\caption{Annual average sunspot groups data for 1989--2004. Notations are the 
same as in Figure \ref{Rich}.}
 \label{Modern1}
\end{figure}

Thus, Figure 4 presents results for the 1989--2004 data set as derived from three 
different temporal averaging routines. Namely, the two top frames (a) show the 
outcome from one-month averaging, the two middle frames (b) present the result 
of two-month averaging, and, finally, the two bottom frames (c) refer to the 
three-month averaging outcome. Data gaps in November 1995, in December 1998 and 
from January till March 2002 are visible. The data show that during the time 
intervals of active Sun, the percentage of anti-Hale groups predominantly does 
not exceed the 7\% level (the dashed horizontal lines on the percentage frames), 
however, during the minimum between the 22nd and 23rd cycles, the value of 
percentage reaches up to 15--25 \% for different averaging intervals. From 
Figure 4 one can conclude that the relative number of anti-Hale groups overcomes 
the 7\%-level when the total number of sunspot groups becomes low (below 
approximately 20, the level marked with the dashed line on the top frame of 
Figure 4).

\begin{figure}
\includegraphics[width=0.50\textwidth]{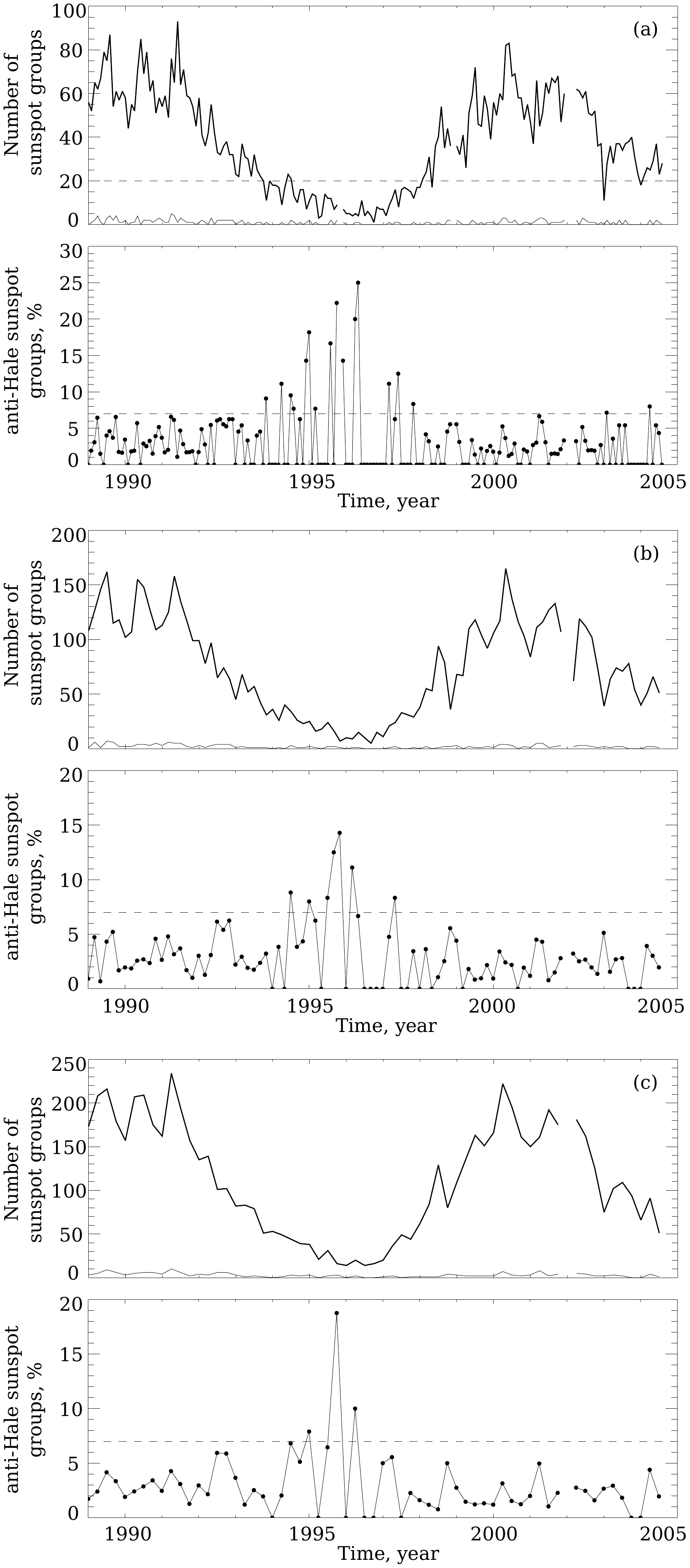}
\caption{Statistics of sunspot groups for 1989--2004: (a) - monthly average; (b) 
- two-month average; (c) - three-month average. For each pair of frames, the top 
frame shows the total number of sunspot groups (thick line), and the number of 
anti-Hale groups (thin line), whereas the bottom frame shows the percentage of 
anti-Hale groups. }
 \label{Modern2}
\end{figure}

We compare the above results with findings of McClintock et al. (2014) who also 
reported an enhancement of percentage of anti-Hale groups near the ends of solar 
activity cycles (Figure 5). Slight differences between our results and that 
reported by McClintock and colleagues are visible, however we presume that they 
might be due to different applied routines to outline a sunspot group. Note, 
that the local peak in the percentage of anti-Hale groups in 1995 is present in 
both studies, compare Figure 3 (bottom frame) and Figure 5 (top frame). 
Thus, an enhancement of the relative number of anti-Hale groups during the solar 
minima can be regarded as a solid result.

\begin{figure}
\includegraphics[width=0.50\textwidth]{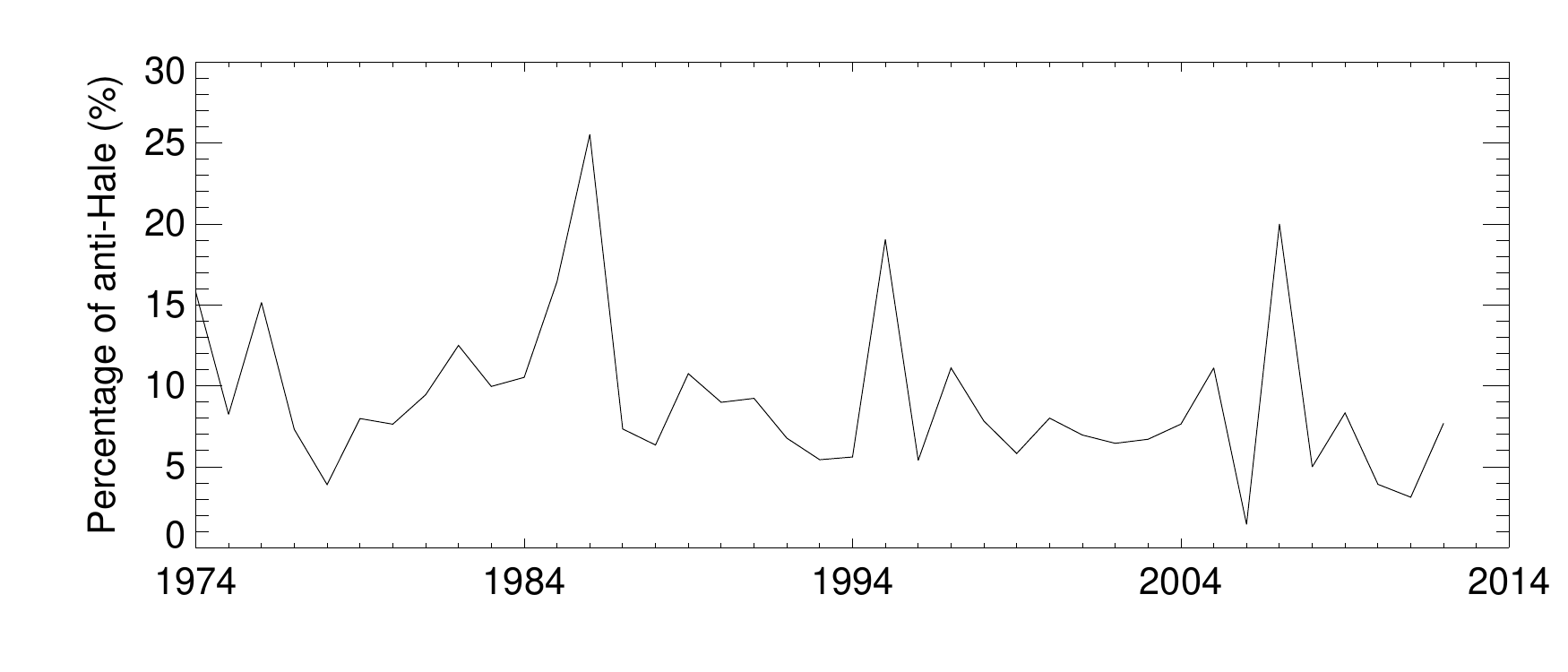}\\
\includegraphics[width=0.50\textwidth]{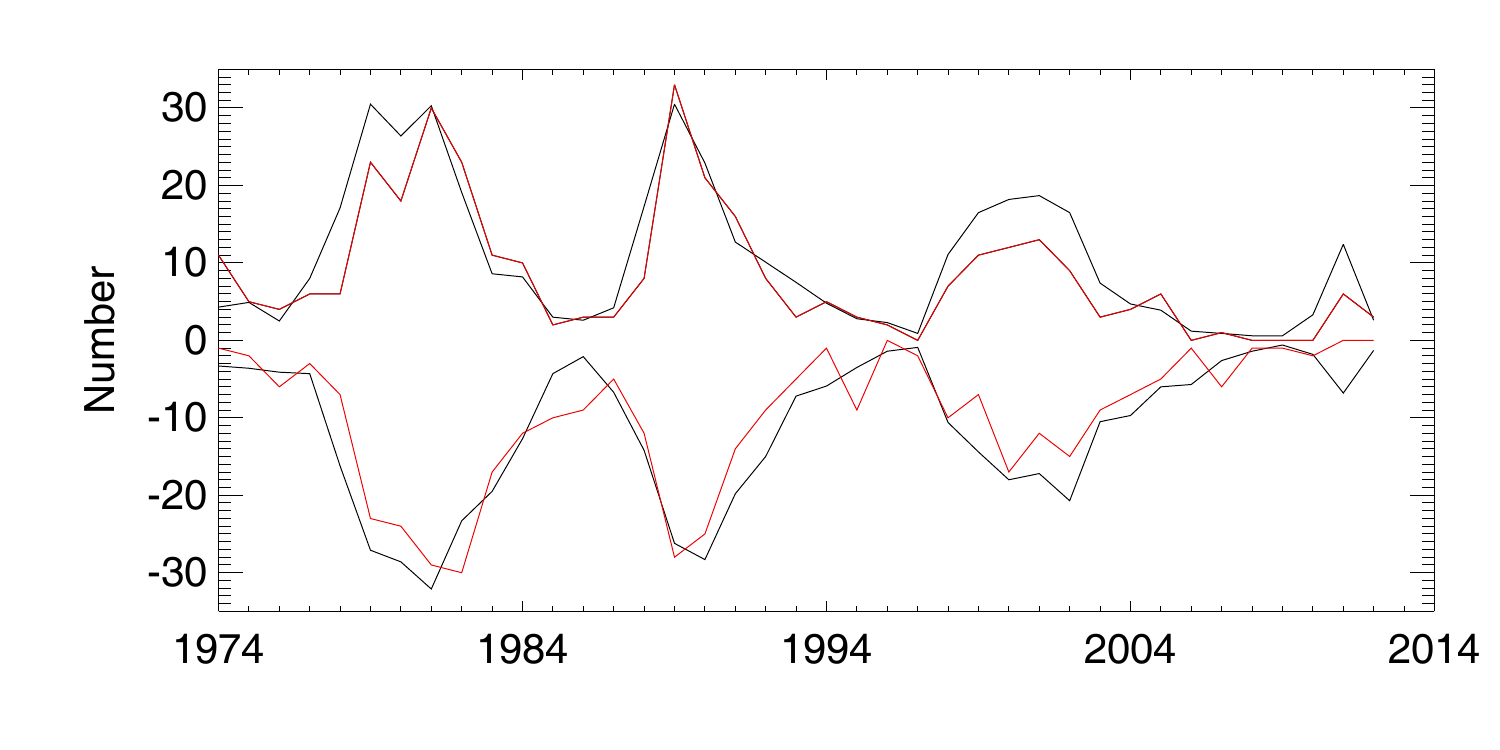}
\caption{ Top: time variations of the percentage of bipolar anti-Hale groups. 
Bottom: time variations of the number of bipolar sunspot regions divided by 10 (black) 
and the number of anti-Hale regions (red). Positive (negative) numbers refer to 
Northern (Southern) hemisphere. Courtesy of McClintock and co-authors (McClintock 
et al., 2014). (A color version of this figure is available in the 
online journal.)}
 \label{MaC}
\end{figure}

\section{Conclusion and discussion}

We demonstrate that the relative number (the percentage) of sunspot groups which 
violate the Hale polarity law (anti-Hale groups) do increase during the minima 
of 11-year solar activity cycles. In accordance with the probabilistic model for 
the magnetic fields, suggested in Section 2, an increase of the relative number 
of anti-Hale groups during the solar minima implies the small-scale dynamo at 
work, because the small-scale dynamo provides a cycle-independent income in the 
number of anti-Hale groups and the total number of groups becomes lower during 
the solar minimum. In other words, the observed statistics of anti-Hale groups 
give a hint that small-scale dynamo is active in the solar interior.

This conclusion is compatible with the inferences made by Stenflo (2012, 2013), 
who considers the small-scale dynamo as a very negligible contributor to the 
total solar magnetic flux because the anti-Hale groups are associated with a 
tiny part of solar magnetic flux only (see Table 1 in Wang \& Sheeley, 1989).

We stress however that a verification of this hint in the framework of other 
approaches remains highly desirable. The point is that the link between 
small-scale magnetic field in solar interior and statistics of anti-Hale groups 
is very far from straightforward, and it is problematic to exclude firmly 
alternative interpretations of the result obtained. In particular, in McClintock 
et al. (2014), it is supposed that the enhancement of the relative number of 
anti-Hale groups could be associated with the solar activity at low latitudes 
via interaction across the equator. We appreciate this option and its importance 
for solar dynamo, however we suppose that it is insufficient to explain the 
observed effect because, during solar minima, anti-Hale groups were observed at 
both low and intermediate/high latitudes (see Figure 1 and time-latitude 
diagrams in Sokoloff \& Khlystova 2010; McClintock et al., 2014).

\section*{Acknowlengements}
D.S. and A.Kh. are grateful for the RFBR financial support under the grant 15-02-01407.
Efforts of V.A. were supported by the Program of the Presidium of Russian Academy of Sciences No. 21.


\end{document}